\documentclass[one column,apsrev,prb,reprint,superscriptaddress]{revtex4-2}
\usepackage{amssymb,amsfonts,amsmath}
\usepackage{adjustbox}
\usepackage{hyperref}
\usepackage{color}
\usepackage{xr}
\usepackage{braket}
\usepackage{xcolor}
\usepackage{float}
\usepackage{setspace}
\usepackage{ulem}
\usepackage{subfigure}

\begin{document}

\title{Optimizing Density Functional Theory for Strain-Dependent Magnetic Properties of MnBi\(_2\)Te\(_4\) with Diffusion Monte Carlo}

\author{Swarnava Ghosh}
\affiliation{National Center for Computational Science, Oak Ridge National Laboratory, Oak Ridge, Tennessee 37831, USA}
\author{Jeonghwan Ahn}
\affiliation{Materials Science and Technology Division, Oak Ridge National Laboratory, Oak Ridge, Tennessee 37831, USA}
\author{Seoung-Hun Kang}
\affiliation{Materials Science and Technology Division, Oak Ridge National Laboratory, Oak Ridge, Tennessee 37831, USA}
\author{Dameul Jeong}
\affiliation{Department of Physics, Kyung Hee University, Seoul, 02447, South Korea}
\author{Markus Eisenbach}
\affiliation{National Center for Computational Science, Oak Ridge National Laboratory, Oak Ridge, Tennessee 37831, USA}
\author{Young-Kyun Kwon}
\affiliation{Department of Physics, Kyung Hee University, Seoul, 02447, South Korea}
\affiliation{Department of Information Display and Research Institute for Basic Sciences, Kyung Hee University, Seoul, 02447, South Korea}
\author{Fernando A. Reboredo}
\affiliation{Materials Science and Technology Division, Oak Ridge National Laboratory, Oak Ridge, Tennessee 37831, USA}
\author{Jaron T. Krogel}
\affiliation{Materials Science and Technology Division, Oak Ridge National Laboratory, Oak Ridge, Tennessee 37831, USA}
\author{Mina Yoon}
\email{myoon@ornl.gov}
\affiliation{Materials Science and Technology Division, Oak Ridge National Laboratory, Oak Ridge, Tennessee 37831, USA}

\begin{abstract}
In this study, we evaluate the predictive power of density functional theory (DFT) for the magnetic properties of MnBi\(_2\)Te\(_4\) (MBT), an intrinsically magnetic topological insulator with potential applications in spintronics and quantum computing. Our theoretical understanding of MBT has been challenged by discrepancies between experimental results and \textit{ab initio} calculations, particularly with respect to its electronic and magnetic properties. Our results show that the magnetic phase diagram of MBT varies significantly depending on the Hubbard $U$ parameter in the DFT framework, highlighting the importance of benchmark calculations. To address these challenges, we establish an optimized Hubbard $U$ approach derived from Diffusion Monte Carlo (DMC) calculations, 
which directly solves the many-body Schr\"{o}dinger equation based on the stochastic process, 
and implement it in the DFT framework. 
Once the optimized $U$ value is determined as a function of strain, we apply it to achieve DMC-level accuracy within our DFT framework. This approach is instrumental in accurately describing the magnetic states of MBT and understanding the underlying mechanisms governing its magnetic properties and their dependence on external factors.
\end{abstract}

\onecolumngrid
Notice: This manuscript has been coauthored by UT-Battelle, LLC, under Contract No. DE-AC0500OR22725 with the U.S. Department of Energy. The United States Government retains and the publisher, by accepting the article for publication, acknowledges that the United States Government retains a non-exclusive, paid-up, irrevocable, world-wide license to publish or reproduce the published form of this manuscript, or allow others to do so, for the United States Government purposes. The Department of Energy will provide public access to these results of federally sponsored research in accordance with the DOE Public Access Plan (http://energy.gov/downloads/doe-public-access-plan).

\maketitle

\section{Introduction}
The discovery of topological insulators - materials that conduct electricity on their surface while acting as insulators inside - opens up new possibilities in condensed matter physics. It allows researchers to explore a wide range of new phenomena, such as the existence of robust, protected edge states that remain insensitive to perturbations~\cite{ zhang2009topological, konig2007quantum, chen2009experimental, hasan2010colloquium, kane2005z, fu2007topological, qi2008topological}. It has also paved the way for the development of innovative electronic devices that promise to be more efficient and versatile than traditional technologies.
Researchers have been exploring the combination of magnetism and topology through techniques such as doping and capping, leading to the creation of a range of fascinating new states with exotic properties~\cite{xiao2018realization, chen2010massive}. One breakthrough in particular has opened up new possibilities in the search for intrinsically magnetic topological insulators. 

The discovery of the MnBi$_2$Te$_4$ (MBT) family has raised the intriguing prospect of materials that exhibit both intrinsic magnetization and topological order~\cite{otrokov2019prediction}. These materials are expected to exhibit remarkable properties such as not only quantum anomalous Hall effect (QAHE)~\cite{chang2013experimental, xiao2018realization, chen2010massive}, but also the quantized magnetoelectric effect, and could hold the key to unlocking a host of new applications in fields such as spintronics and quantum computing~\cite{li2019intrinsic, he2020mnbi2te4, zhang2019topological, deng2020quantum, qi2008topological}.
MBT is a van der Waals material that consists of septuple layers (SLs) arranged as Te-Bi-Te-Mn-Te-Bi-Te. These materials exhibit magnetic and topologically nontrivial phases, with MBT demonstrating a large magnetization arising from the magnetic order of Mn atoms. The magnetic properties of MBT vary depending on the layer number, with even-numbered layers exhibiting antiferromagnetism and odd-numbered layers showing ferromagnetism~\cite{ li2019intrinsic, he2020mnbi2te4}. These magnetizations can be controlled by external factors such as temperature and strain~\cite{klimovskikh2020tunable, yan2021elusive, guo2021pressure, ovchinnikov2021intertwined,xue2020control, gao2021mbt, otrokov2019prediction, lei2021metamagnetism}. 

Although MBT has been extensively studied, there remain significant discrepancies between experimental results and theoretical calculations that prevent a complete understanding of its properties~\cite{yan2021elusive, ahn2023mbt}.  For example, theoretical predictions suggest that the surface (001) of the antiferromagnetic MBT should exhibit a band gap due to broken space inversion symmetry, but experimental results often show that the band gap remains closed~\cite{ li2019dirac, hao2019gapless, chen2019topological, swatek2020gapless, nevola2020coexistence}. 
Recent studies suggest that stacking faults on the MBT surface may be a critical factor in these discrepancies~\cite{ahn2023mbt}. These faults cause deviations from the ideal structure, leading to nearly gapless surface states because of charge redistribution and weakened interlayer coupling. This highlights a gap in our theoretical understanding of the electronic properties of the MBT and emphasizes the need for a comprehensive approach to reconcile theoretical predictions with experimental observations. 

To address these discrepancies, we investigate the strain-dependent magnetic properties of MBT. Strain variation from the ideal structure can significantly affect the magnetic properties, as previous studies have shown that strain can alter the magnetic phase transitions and enhance the magnetic moments of materials. By understanding the effects of strain on magnetism, we can better understand the critical behavior resulting from the correlation between structure (strain) and magnetism. 

By employing first-principles approaches, such as diffusion Monte Carlo\cite{ceperley1980ground} (DMC) and density functional theory (DFT), researchers can obtain valuable insights into the electronic and magnetic properties of MBT under different conditions, such as defects and strain. The Hubbard $U$ parameter is introduced in the DFT scheme to capture electron correlation effects, which are important for accurately describing the magnetic properties. 
Many computational studies on MBT have used different Hubbard $U$ values, such as 3~eV~\cite{zhu2021tunable}, 4~eV~\cite{lai2021defect, du2021tuning, xue2020control}, 5~eV\cite{hao2019gapless}, 5.34~eV~\cite{otrokov2019prediction, klimovskikh2020tunable}, resulting in diverse results depending on these parameters. Obtaining accurate magnetic properties of MBT is of high interest, as it is directly related to its topological properties as well~\cite{luo2023mbt}. 

It is essential to optimize the DFT framework by calibrating the Hubbard $U$ parameter to capture the electron correlation from the magnetic atoms.  This optimization takes into account the dependence of the exchange-correlation (XC) functional and the sensitivity of $U$ on magnetic and structural properties not only in the MBT but also in other similar systems. Accurate modeling of these dependencies is crucial for reconciling theoretical predictions with experimental results. Furthermore, this approach demonstrates the need for DMC benchmarks to validate and refine the Hubbard $U$ values, thus ensuring the reliability of the DFT framework in describing the magnetic states under varying strains.

In this paper, we study the magnetic phase transitions of MBT under strain using DFT + $U$ and DMC approaches. Using DMC calculations as benchmark, we determined the optimal $U$ values based on the applied strain, ensuring unique and optimized Hubbard $U$ values. We then used these optimized $U$ values in the DFT to calculate the energy difference between different magnetic configurations under strain. Our results show that the magnetic phase diagram of MBT varies significantly depending on the $U$ parameter, highlighting the importance of benchmark calculations and optimization of $U$ values in DFT. Once the optimized $U$ value is determined as a function of strain, we apply it to achieve DMC-level accuracy within our DFT framework. These studies enhance our understanding of the underlying mechanisms governing the observed magnetic properties and their dependence on external factors.

\section{Computational Approaches}
\subsection{Density Functional Theory}
We utilized the Vienna {\it{ ab initio}} simulation package (VASP) based on density functional theory to carry out a first-principles study and compute the magnetic properties of MBT~\cite{{hohenberg1964inhomogeneous,kohn1965self},kresse1993ab,kresse1996efficient}. To ensure convergence, we adopted a plane wave cutoff of 400~eV with projector augmented wave method (PAW) pseudopotentials~\cite{blochl1994projector} and employed an $18\times18\times1$ $\Gamma$-centered $k$-points sampling with a convergence criterion of 0.05~meV in magnetic anisotropy energy for pristine MBT. Since the structure considering the antisite defect is a $3\times3\times1$ supercell of the pristine unit cell, the k-points sampling was also set to $6\times6\times1$ accordingly. 

As exchange-correlation functional, generalized gradient approximation (GGA) of Perdew-Burke-Ernzerhof (PBE)~\cite{GGA}, and Ceperley-Alder form~\cite{ceperley1980ground} within the local density approximation (LDA) was used. We set the Hellman-Feynman force criterion to 0.01~eV/$\AA$ for structural and atomic relaxation and included spin-orbit coupling (SOC) interaction in all calculations. Mn has partially filled $d$-orbitals, and their electronic behavior can be intricate, involving strong electron-electron interactions. Hence, in such cases, the introduction of Hubbard $U$ in the framework of DFT is necessary to account for the on-site Coulomb repulsion between electrons in the transition metal $d$-orbitals. We studied the magnetic properties of MBT using the $U$ parameter values between 0~eV and 4.4~eV on Mn 3$d$ orbitals, and  the DMC optimized values for the given strain were used to evaluate the accurate magnetic anisotropy energy.

For selected cases, we compared the electronic ground states using the all-electron method as implemented in Elk~\cite{elk} and the pseudopotential method as implemented in VASP~\cite{kresse1993ab, kresse1996efficient}. In Elk, the DFT+$U$ method is implemented using either the around mean-field (AMF) \cite{amf} or fully localized limit (FLL) \cite{fll} schemes. This comparison allows us to evaluate the impact of different DFT+$U$ implementations on the magnetic properties of MBT, providing a more comprehensive understanding of the impact of different approximations of the theory under different computational approaches. Readers are referred to the SI for for details and discussion of the all-electron calculations.

\subsection{Diffusion Monte Carlo}
DMC is a real-space method to directly solve the many-body Schr\"{o}dinger equation based on the stochastic process. 
DMC calculations were implemented with QMCPACK~\cite{kim2018,kent2020} code under the fixed-node approximation~\cite{anderson1975,anderson1976} that makes the projected state remain antisymmetric as a fermionic system.
Because the Hubbard $U$ can be served as a variational parameter in optimizing the nodal surface used in the fixed-node DMC calculations, one can find an optimal value of Hubbard $U$ to yield the lowest DMC energy for the given structure.
This allows us to model values of Hubbard $U$ depending on the strain of the system, as discussed later. 
We employed the Slater-Jastrow type function for a many-body trial wave function. The Slater determinants were constructed with DFT Kohn-Sham~\cite{kohn1965self} orbitals based on the PBE+$U$ functional
where the corresponding DFT calculations were performed with QUANTUM ESPRESSO~\cite{gianozzi2009} package. 

We considered one-, two- and three-body Jastrow factors to describe electron-ion, electron-electron, and electron-electron-ion correlation effects, respectively. 
The norm-conserving pseudopotentials developed within ccECP approach~\cite{Bennett2017,Bennett2018} were used to represent the atomic cores as in our previous DMC-informed DFT studies for the MBT system~\cite{bennett2022magnetic}.
The Jastrow optimizations were done with variational Monte Carlo calculations based on the linear method of Umrigar {\it et al.}~\cite{umrigar2007} and subsequent DMC calculations were done using a time step of $\tau = 0.005$ Ha$^{-1}$. To evaluate non-local pseudopotentials, imagine time projections were treated with size-consistent T-moves~\cite{Casula2010}. 

\section{Results}
\subsection{Hubbard U-value dependent magnetic states}
%
\begin{figure}[t]
\includegraphics[width=0.7\linewidth, angle=0] {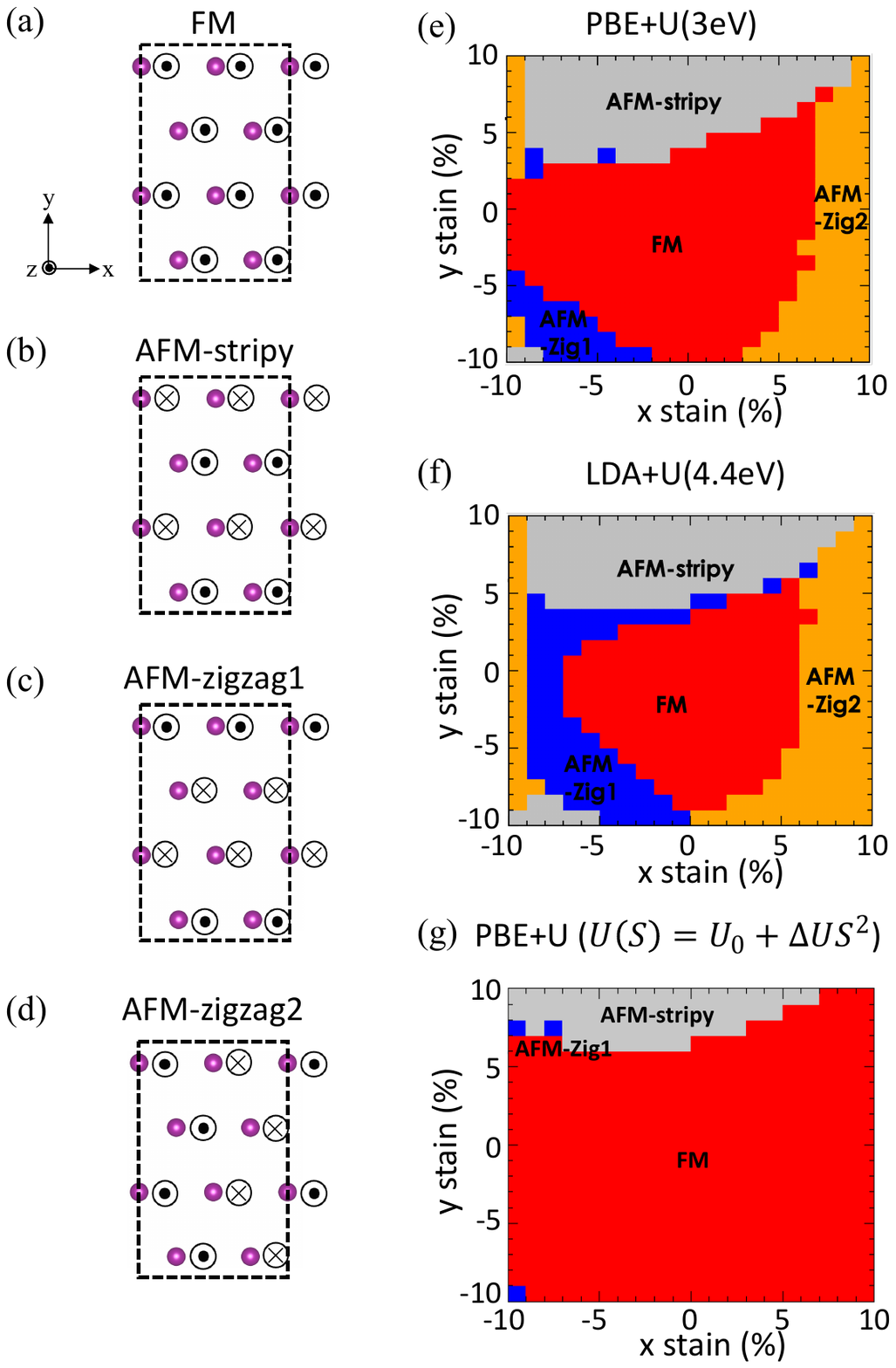}
\caption{Magnetic configurations of the 1 septuple layer (SL) MBT considered in this study: FM (a), AFM-stripy (b), AFM-zigzag 1 (c), and AFM-zigzag 2 The stable magnetic phases among these four magnetic states depending on the in-planar strain, x and y, up to 10\% of the tensile and compressive strains. The magnetic phase diagram changes drastically depending on the choice of exchange-correlation functional and Hubbard $U$ values: (e) PBE+$U$ (=3~eV), (f) LDA+$U$ (=4~eV), and (g) PBE+$U_{DMC}$, DMC optimized $U$ depending on strain.}
\label{mag_phase_diagram}
\end{figure} 
%
Using the DFT+U approach we study the influence of the Hubbard $U$ value  on the magnetic ground state in 1 SL  MBT. Figure~\ref{mag_phase_diagram} illustrates the magnetic configurations of the 1 SL MBT considered in this study - they are ferromagnetic (FM) and three different antiferromagnetic (AFM) configurations forming a stripe pattern, and two different types of zigzag ordering of spin up and down components, either along $x$ or $y$ directions. The stable magnetic phases among these four magnetic states depend on the in-planar strain (on the $xy$ plane), up to 10\% of the tensile and compressive strains. Most first-principles studies of the MBT have taken a liberty in choosing Hubbard $U$ values between $\approx$ 3 eV and $\approx$ 5 eV~\cite{zhu2021tunable, lai2021defect, du2021tuning, xue2020control, hao2019gapless, otrokov2019prediction, klimovskikh2020tunable}. 

Our calculations show that the choice of the Hubbard $U$ value is very critical, as the magnetic phase diagram can look very different depending on the DFT parameters, such as the exchange correlation functional (local density approximation (LDA) vs. generalized gradient approximation (GGA)) and Hubbard $U$ values. 
In the absence of strain, DMC predicts a value of 4.0(2) eV for the monolayer and 4.4 eV for LDA+$U$ of the MBT bulk, while PBE+$U$ gives a value of 3.5 eV for the bulk. If we keep these values independent of strain, the magnetic phase diagram varies significantly depending on the DFT parameters. For example, PBE+$U$ (=3 eV) calculations predict that FM states are favorable over a wide range of $x$-strain, while LDA+$U$ (=4.4 eV) calculations show a much larger range of AFM zigzag1 phases and a smaller range of FM phases in the strain range (see Fig.~\ref{mag_phase_diagram}). The DMC optimized $U$ values, $U_{DMC}$, which will be discussed later, predict something completely different (see Fig.~\ref{mag_phase_diagram} (g)).

\begin{figure}[h]
\subfigure[GGA+U]{\includegraphics[width=0.7\linewidth, angle=0] {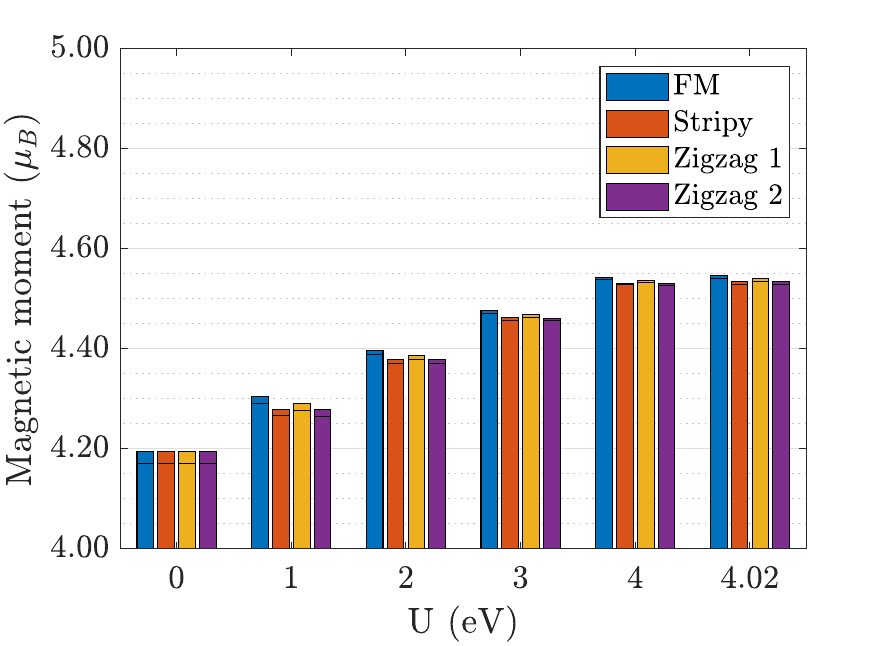}}
\subfigure[LDA+U]{\includegraphics[width=0.7\linewidth, angle=0]{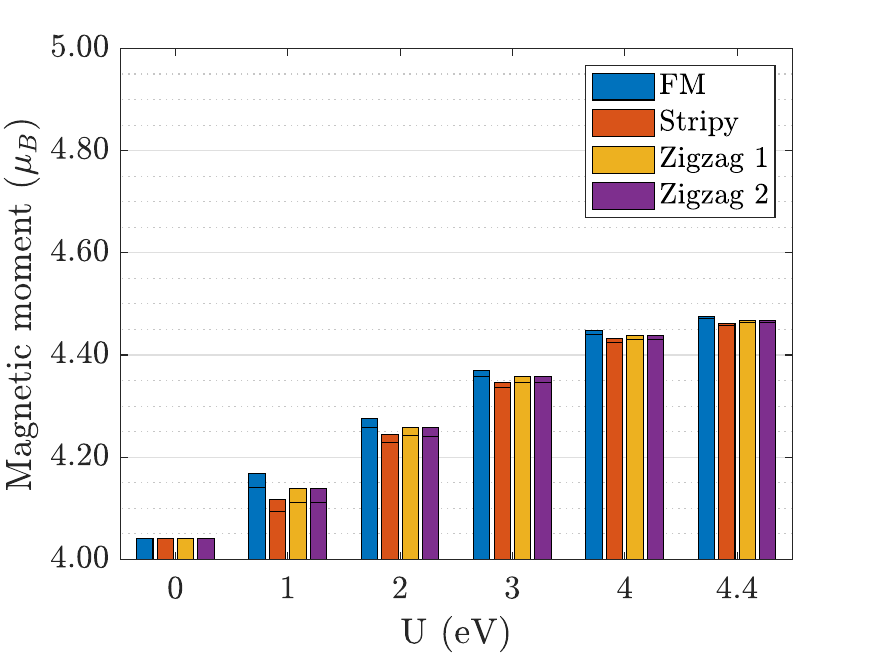}}
\caption{Magnetic moments of the Mn atoms in the pristine (unstrained) 1 SL MBT based on the calculations with a) GGA+$U$ and b) LDA+$U$.} 
\label{moment_U}
\end{figure}

On the other hand, for the given magnetic configurations, either FM or different AFM states, the magnetic moments are not so sensitively changed by $U$ values, and above ~3eV the values more or less converge. Figure \ref{moment_U} shows the magnetic moments at Mn atoms for different values of $U$ for a) GGA+$U$ and b) LDA+$U$ cases. The value of the magnetic moment increases from 4.2 $\mu_B$ to 4.55 $\mu_B$ as $U$ is increased from 0 to 4.02 eV in the GGA+$U$ case and from 4.05 $\mu_B$n to 4.45 $\mu_B$ as $U$ is increased from 0 eV to 4.4 eV in the LDA+$U$ case. Spin-orbit coupling and magnetic configuration (FM, stripy, zigzag 1, and zigzag 2) have a negligible effect on the magnetic moment.

%
%

%
\begin{figure}[h]
\begin{center}
\subfigure[GGA+U]{\includegraphics[width=0.7\linewidth, angle=0] {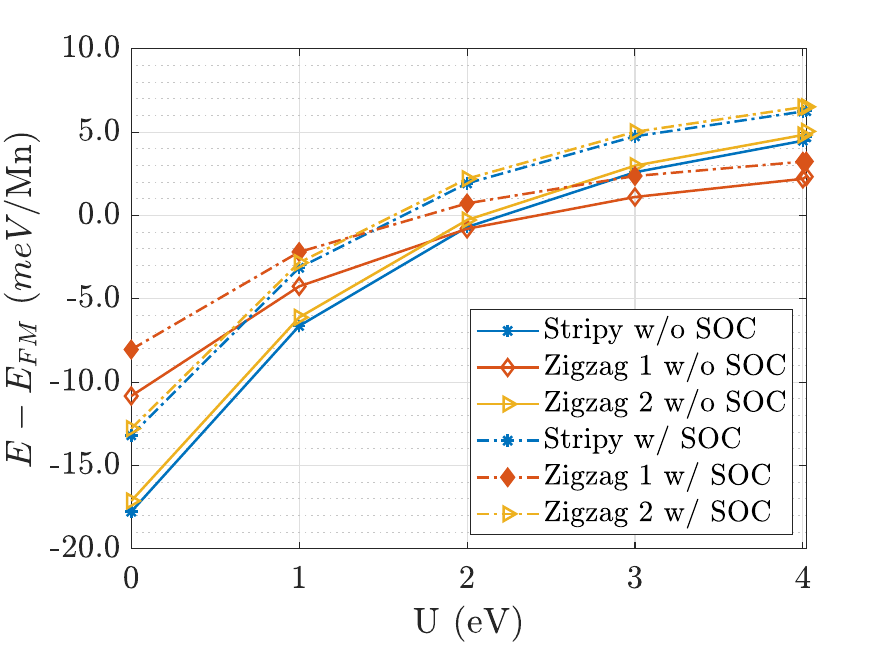}}
\subfigure[LDA+U]{\includegraphics[width=0.7\linewidth, angle=0] {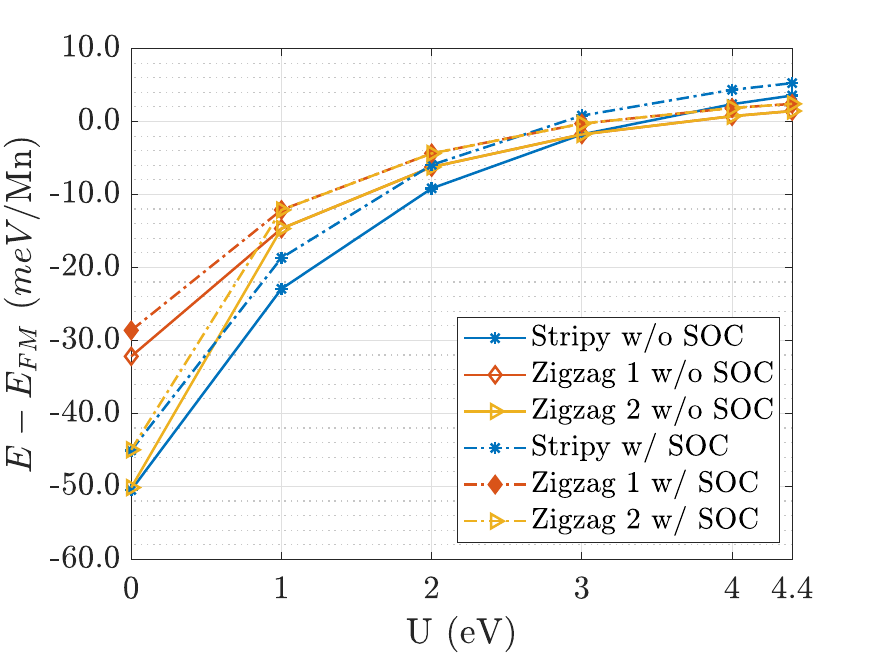}}
\caption{ Energy difference versus different choices of the $U$ parameter for a) GGA+$U$ and b) LDA+$U$ exchange correlation functionals. The energy difference is defined as the difference between the different magnetic states and the ferromagnetic state. } 
\label{energy_U}
\end{center}
\end{figure} 
%
Next, we study the influence of the Hubbard $U$ on the energy difference between different magnetic states (see Fig. \ref{mag_phase_diagram}). The supercell has been optimized for each value of $U$. Fig. \ref{energy_U} shows the energy difference as a function of $U$ for the GGA+$U$ and LDA+$U$ calculations. The energy difference is defined as the difference between the magnetic state and the ferromagnetic state ($E-E_{FM}$). From Fig. \ref{energy_U}a we observe that for $U$=0 eV, AFM stripy is the favored magnetic ground state for both GGA and LDA exchange correlations. When the value of $U$ is further increased to 4.02 eV for GGA+$U$ and 4.4 eV for LDA+$U$, the favored magnetic ground state changes to an FM state. The transformation from AFM stripe to FM occurs between $U$=2-3 eV for GGA+$U$ and 3-4 eV for LDA+$U$. Note also that the relative ordering of the magnetic states is different for the GGA+$U$ and LDA+$U$ cases. For GGA + $U$=4.02 eV, $E_{FM} < E_{zigzag 1} < E_{striped} < E_{zigzag 2} $, and for LDA + $U$=4.4 eV, $E_{FM} < E_{zigzag 1} = E_{zigzag 2} < E_{stripy} $. 

We conclude that a too small $U$ value leads to an underestimation of the magnetic moment, which is known from neutron scattering experiments to be about 4.7(1) $\mu_B$~\cite{Ding_PhysRevB_2020} in bulk MBT. 
In addition, a too small U value erroneously favors non-ferromagnetic states and possibly purely paramagnetic states at zero strain. Most of the differences between LDA+$U$ and PBE+$U$ can be explained by a 1-1.5 eV shift in the $U$ value. This demonstrates the critical importance of accurately determining the Hubbard $U$ parameter to correctly describe the magnetic properties and phase transitions of MBT under various conditions. The optimized $U$ values derived from DMC calculations provide a framework for achieving DMC-level accuracy within the DFT framework, ensuring meaningful and reliable results. 

Our comparative analysis using both pseudopotential and all-electron calculations highlights the influence of DFT pseudopotential errors, which can lead to $\lesssim$ 10\% errors in the magnetic moments. In addition to the choice of U-value and its implementation method (AMF vs. FLL), the magnetic properties could be significantly affected, with all-electron calculations providing a more detailed insight compared to pseudopotential methods (see SI for details).

\subsection{Determining optimized-$U$ depending on strain via DMC}
\label{sec:dmc_u_model}
\begin{figure}[h]
\centering
\includegraphics[width=0.7\linewidth] {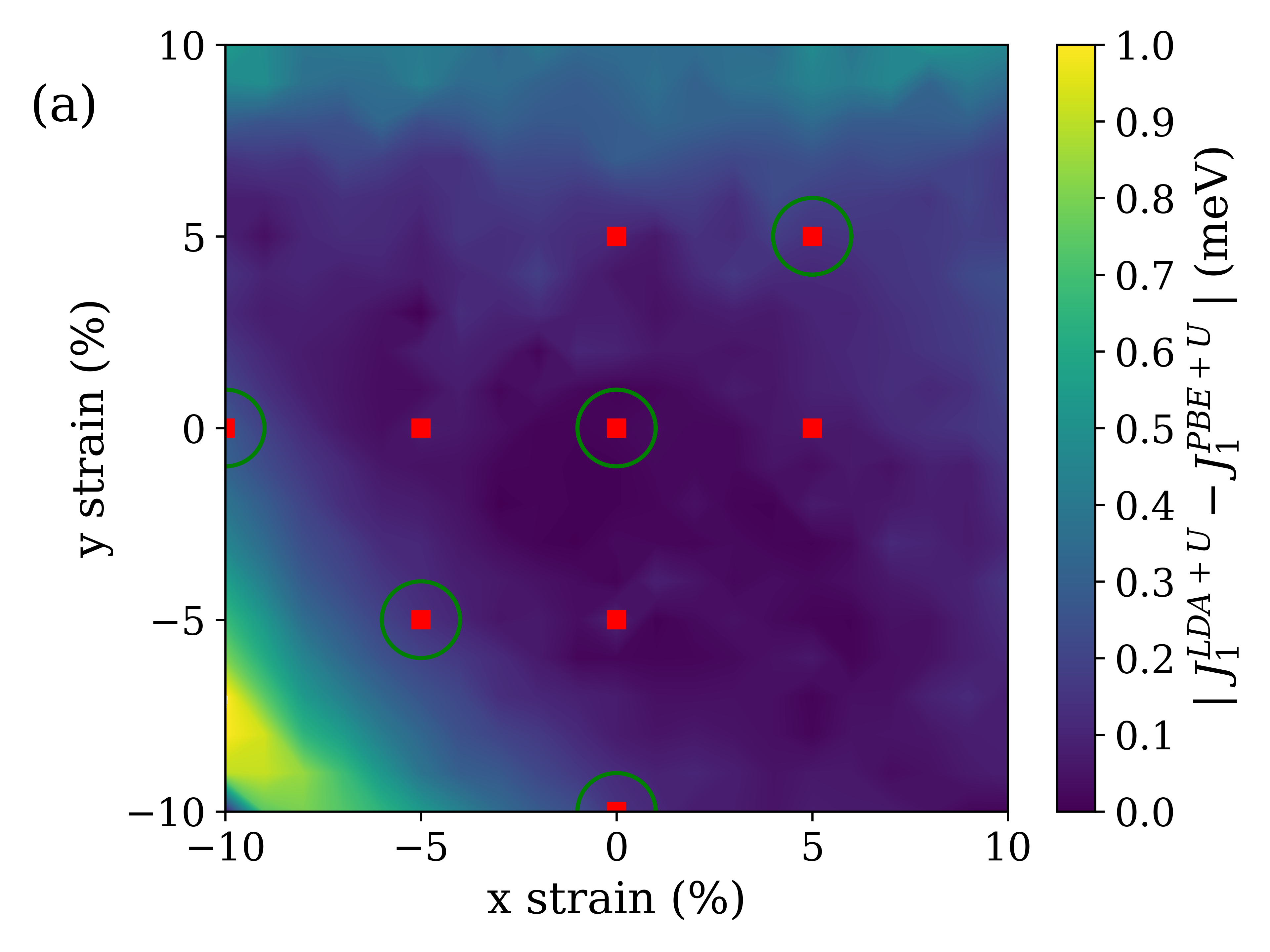}
\includegraphics[width=0.7\linewidth] {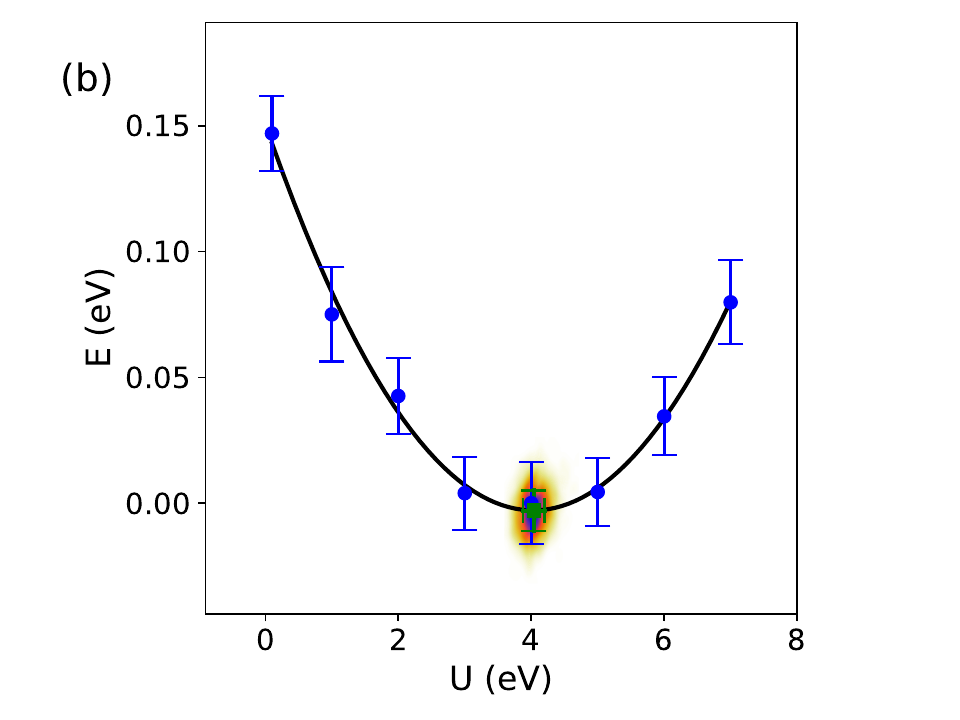}
\caption{Sensitivity analysis to guide optimal determination of Hubbard U via DMC.  In (a) the absolute difference in $J_1$ exchange parameter between LDA+$U$=4.4 eV and PBE+$U$=3 eV is used to indicate the spin model sensitivity to U.  Strain regions showing high sensitivity (large absolute $J_1$ difference) are used to select $x$ and $y$ strain points (circled in green) for subsequent determination of optimal $U$ via DMC.  In (b) the DMC optimized $U$ (for PBE+$U$) is determined by the variational principle at zero strain.  The distribution of minima resulting from resampled quadratic fits are shown in color. } 
\label{fig:dmc_design}
\end{figure}

Because strain is expected to alter the magnetic properties of MBT, as mentioned above, 
it is highly beneficial to understand a relationship between strain and Hubbard $U$.  
For this purpose, we use highly accurate diffusion Monte Carlo (DMC) methods to determine 
the optimal $U$ for PBE+$U$ and how it varies with strain in the 1 SL MBT.
Since DMC is computationally demanding, we first performed a sensitivity analysis based 
on the dominant exchange contribution ($J_1$) that enters into spin models for the magnetic 
behavior of this material.  As a starting point, we evaluated $J_1$ vs strain with Hubbard $U$ functionals selected from previous studies.  A prior DFT study has used PBE+$U$=3 eV in MBT bulk and another prior DMC study found $U$=4.4 eV as optimal for the same system and so we obtained $J_1$ vs strain for both of these functionals.  In Fig. \ref{fig:dmc_design}a we show the absolute difference in $J_1$ calculated with these two functionals as a function of strain for the 1SL MBT.  Large absolute differences indicate strain regions that are sensitive to the underlying assumptions of the DFT functionals, and thus serve to idenify where better information about the functional is most needed.  

Based on this, we selected four strain points (highlighted by green circles in Fig. \ref{fig:dmc_design}a), in addition to the origin, to perform nodal optimization as a function of $U$ for PBE+$U$ with DMC.  These points were $(0 \%,0\%),~(5 \%,5\%),~(-5 \%,-5\%),\\
~(-10 \%,0\%),~(0 \%,-10\%)$ x,y strain. The resulting nodal optimization for zero strain is shown in Fig. \ref{fig:dmc_design}b.
Quadratic fits were performed based on resampled DMC energy data to produce a best estimate, including statistical uncertainty for the optimal $U$.  The independent estimate for $U$ at zero strain is $4.04(16)$ eV, which may differ slightly from the previously determined optimal $U$ value of 3.5(2) eV for PBE+$U$ in MBT bulk\cite{bennett2022magnetic}.

\begin{figure}[h]
\centering
\includegraphics[width=0.7\linewidth] {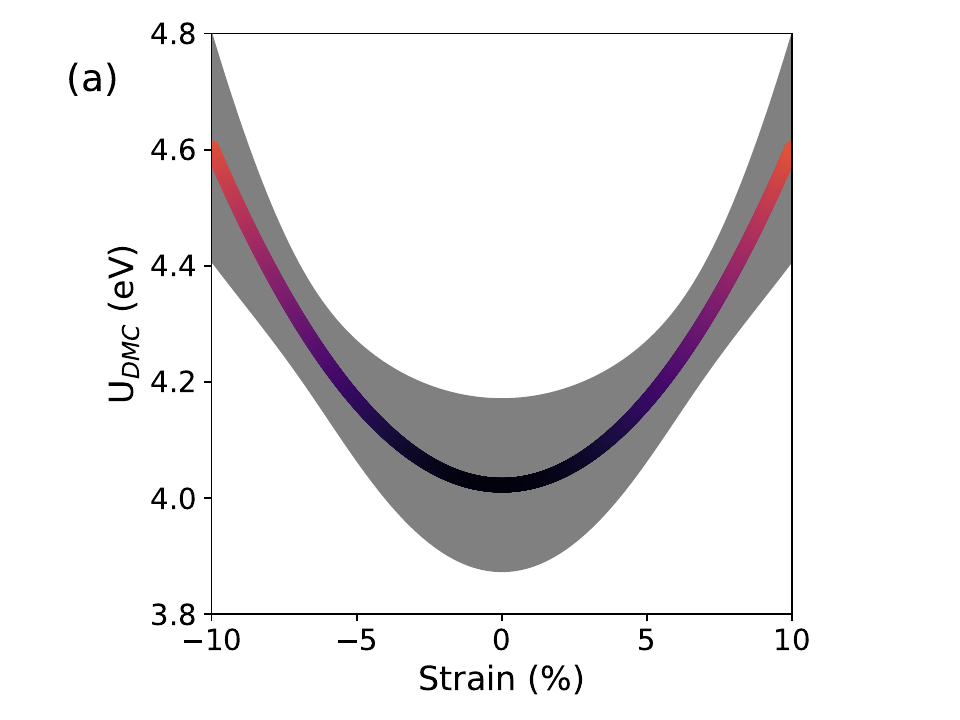}
\includegraphics[width=0.7\linewidth] {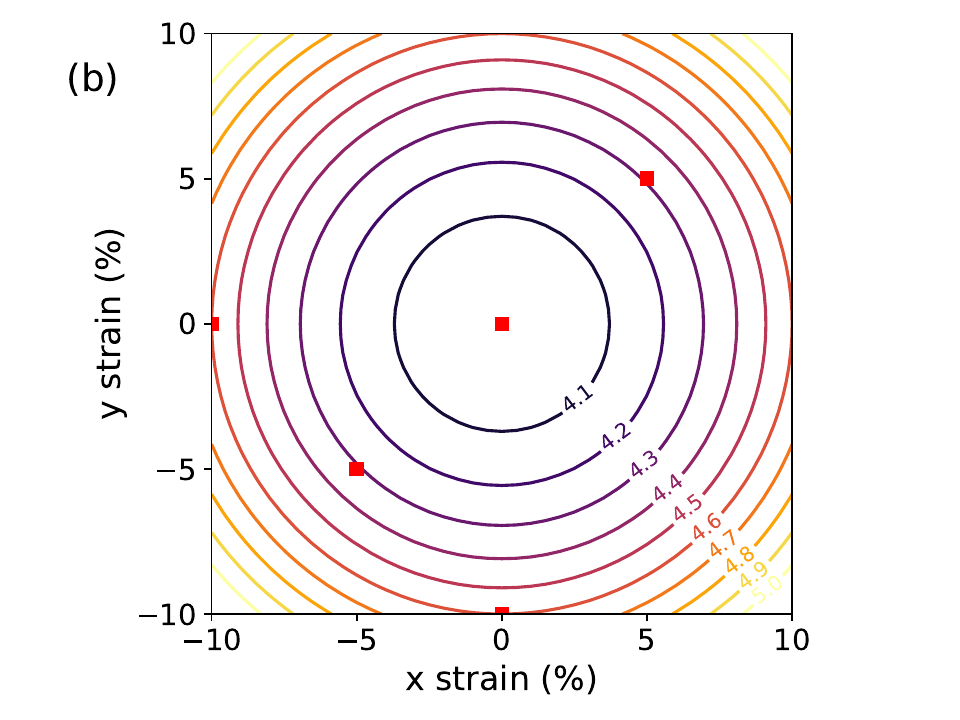}
\caption{Variation in DMC optimized U vs strain.  Panel (a) shows the near isotropic strain model for U supported by the DMC data collected at multiple strains ($U_{\mathrm{DMC}}(S)=U_0+\Delta U S^2$), with $S$ being the absolute magnitude of total strain.  Panel (b) shows the same model for U, but as contours over the x/y strain field of the 1SL MBT.} 
\label{fig:dmc_U_model}
\end{figure}

The resulting optimal U values for all strain points (Fig.~\ref{fig:dmc_U_model}) were then combined to arrive at a simple model for the dependence of $U$ on strain for 1 SL MBT.  For all values of strain, we found an increase in the optimal $U$.  The simplest model supported by these data with analyticity at zero strain is a simple quadratic form.  Since the difference in optimal $U$ values between points with identical total strain $S=\lvert\lvert\vec{S}\rvert\rvert$ was statistically indistinguisable from zero, an isotropic model was used:
\begin{equation}
U_\mathrm{DMC}(S) = U_{0}+\Delta{U}S^2,
\label{udmc}
\end{equation}
Based on this model, we found fitting parameters of $U_0=4.02(15) eV$ and $\Delta U=0.0058(29)$ eV.  The resulting model, with uncertainty shown as a grey background, appears in Fig. \ref{fig:dmc_U_model}a.  The same model is shown over the 2D strain field in Fig. \ref{fig:dmc_U_model}b with contours showing the U values for each value of strain.  Although the model used here is isotropic, we point out that its impact on the magnetic properties is not.  Base on the sensitivity analysis performed as shown in Fig. \ref{fig:dmc_design}, its main impact is to alter $J_1$ along the biaxial compressive and uniaxial compressive (y-axis) strain directions.

%
%

%
%

\section{Conclusion}
In summary, we studied the magnetic properties of MBT under various strains using DFT and DMC. Our results show that strain significantly affects the magnetic phase transitions of MBT, with the magnetic phase diagram varying significantly depending on the Hubbard $U$ parameter. This highlights the importance of benchmark calculations and optimization of $U$ values within the DFT framework to achieve accurate descriptions of magnetic states. By implementing the optimized $U$ values derived from the many-body DMC calculations, we have ensured DMC-level accuracy within the DFT framework, providing a robust and reliable approach for studying the magnetic properties of MBT. These results contribute to a better understanding of MBT and provide a methodological framework that may be applicable to other magnetic topological insulators.
\clearpage

\section*{Acknowledgement}
This work was supported by the U.S. Department of Energy, Office of Science, Office of Basic Energy Sciences, Materials Sciences and Engineering Division (F.A.R., M.E., M.Y., S.G.) and by the U.S. Department of Energy (DOE), Office of Science, National Quantum Information Science Research Centers, Quantum Science Center (S.-H.K.). 
J.A. (DMC calculations, writing) and J.T.K. (mentorship, analysis, writing) were supported by the U.S. Department of Energy,
Office of Science, Basic Energy Sciences, Materials Sciences and Engineering Division, as part of the Computational Materials Sciences Program and Center for Predictive Simulation of Functional Materials
This work (calculating magnetic anisotropy) was also partly supported by the Korean government (MSIT) through the National Research Foundation of Korea (NRF) (2022R1A2C1005505 and 2022M3F3A2A01073562) and Institute for Information \& Communications Technology Planning \& Evaluation (IITP) (2021-0-01580) (D.J., Y.-K.K.).  An award of computer time was provided by the Innovative and Novel Computational
Impact on Theory and Experiment (INCITE) program. This research used resources of the Oak Ridge Leadership Computing Facility at the Oak Ridge National Laboratory, which is supported by the Office of Science of the U.S. Department of Energy under Contract No. DE-AC05-00OR22725.  This research also used resources of the National Energy Research Scientific Computing Center, a DOE Office of Science User Facility supported by the Office of Science of the U.S. Department of Energy under Contract No. DE-AC02-05CH11231 (M.E and S.G) and using NERSC award BES-ERCAP0024568. 

\bibliographystyle{apsrev}
\bibliography{biblio.bib}

\end{document}


\title{Optimizing Density Functional Theory for Strain-Dependent Magnetic Properties of MnBi\(_2\)Te\(_4\) with Diffusion Monte Carlo}

\author{Swarnava Ghosh}
\affiliation{National Center for Computational Science, Oak Ridge National Laboratory, Oak Ridge, Tennessee 37831, USA}
\author{Jeonghwan Ahn}
\affiliation{Materials Science and Technology Division, Oak Ridge National Laboratory, Oak Ridge, Tennessee 37831, USA}
\author{Seoung-Hun Kang}
\affiliation{Materials Science and Technology Division, Oak Ridge National Laboratory, Oak Ridge, Tennessee 37831, USA}
\author{Dameul Jeong}
\affiliation{Department of Physics, Kyung Hee University, Seoul, 02447, South Korea}
\author{Markus Eisenbach}
\affiliation{National Center for Computational Science, Oak Ridge National Laboratory, Oak Ridge, Tennessee 37831, USA}
\author{Young-Kyun Kwon}
\affiliation{Department of Physics, Kyung Hee University, Seoul, 02447, South Korea}
\affiliation{Department of Information Display and Research Institute for Basic Sciences, Kyung Hee University, Seoul, 02447, South Korea}
\author{Fernando A. Reboredo}
\affiliation{Materials Science and Technology Division, Oak Ridge National Laboratory, Oak Ridge, Tennessee 37831, USA}
\author{Jaron T. Krogel}
\affiliation{Materials Science and Technology Division, Oak Ridge National Laboratory, Oak Ridge, Tennessee 37831, USA}
\author{Mina Yoon}
\email{myoon@ornl.gov}
\affiliation{Materials Science and Technology Division, Oak Ridge National Laboratory, Oak Ridge, Tennessee 37831, USA}

\date{\today}

\maketitle

\newpage

\section{Magnetic states depending on core electrons - pseudopotential vs. all-electron calculations}

We performed all-electron calculations using Linearized Augmented Plane Wave (LAPW) basis with local orbitals as implemented in the  Elk \cite{elk} simulation package to carry out the first-principles study and compute the magnetic properties of MBT. We employed $12\times8\times1$ $\Gamma$-centered $k$-points sampling. The generalized gradient approximation (GGA) of Perdew-Burke-Ernzerhof (PBE)~\cite{GGA} exchange-correlation functional was used. Depending on the choice of the magnetic state of the Mn atoms, the magnetic moment direction was fixed but its magnitude was allowed to evolve. The angular momentum cutoff for the APW functions was chosen as 10 and 10 extra states were chosen per atom and spin.

A unit cell containing four manganese atoms (i.e. $2 \times 2 \times 1$) is considered for our calculations. Three cases of magnetism are considered. First is a ferromagnetic state (FM), where all the magnetic moments on the Mn atoms point in the same direction. Second, an intermediate Ferrimagnetic (FiM) case, where one of the Mn atoms has a moment in the opposite direction. Third, an antiferromagnetic (AFM) case where two moments point up and two moments point down. For each of these cases, we calculate the electronic ground state using the all-electron method as implemented in Elk \cite{elk}. In Elk, the DFT+$U$ method is implemented around a mean-field (AMF) \cite{amf} or fully localized limit (FLL)  \cite{fll}. 

In the AMF implementation, the total energy and the effective potential are corrected by
\begin{eqnarray}
    \Delta E _{AMF} &=& - \frac{U-J}{2} \sum_{\sigma} \Tr (\delta \rho_{mm'}^{\sigma} \cdot \delta \rho_{mm'}^{\sigma}) \,\,, \\
     \Delta V _{AMF} (mm'\sigma) &=& -(U-J) (\rho_{mm'}^{\sigma} - n_{\sigma} \delta_{mm'}) \,\,.
\end{eqnarray}
where $U$ and $J$ are the spherically averaged Hubbard repulsion term and the intra-atomic exchange term for electrons with angular momentum $l$, $\sigma=\pm 1$ is used for indexing the electron spin, $\rho_{mm'}^{\sigma}$ is the orbital occupation matrix, $\delta\rho^{\sigma}_{mm'} = \rho_{mm'}^{\sigma} - n_{\sigma} \delta_{mm'}$ and $n_{\sigma} = \Tr(\rho^{\sigma})/(2l+1)$. The AMF approximation assumes the limit of uniform occupancy and the mean field is defined such that the correction to the electronic potential averaged over all the occupied states sum up to zero \cite{Petukhov}. For strongly correlated systems, this approximation can lead to erroneous results.

In the FLL implementation, the corrections to the total energy and the effective potential is
\begin{eqnarray}
    \Delta E _{FLL} &=& - \frac{U-J}{2} \sum_{\sigma} [ \Tr  (\delta \rho_{mm'}^{\sigma} \cdot \delta \rho_{mm'}^{\sigma}) - (2l+1) n_{\sigma}] \,\,, \\
     \Delta V _{FLL} (mm'\sigma) &=& -(U-J) \left(\rho_{mm'}^{\sigma} - \frac{1}{2} \delta_{mm'}\right) \,\,.
\end{eqnarray}
The AMF implementation of the DFT+U method gives a positive correction to the total energy whereas the FLL implementation gives a negative correction to the total energy. Note that the "true" correction should be zero. This can be done by a linear combination of the FLL and AMF corrections\cite{Petukhov}.

The value of magnetic moments on the Mn atom is also significantly influenced by the choice of $U$ and the nature of its application. When U=0 eV, the magnetic moment is between 4.19-4.25 Bohr magneton ($\mu_B$). When $U$ = 3~eV, the magnetic moment is between 4.44 - 4.92 $\mu_B$ using the FLL implementation and 4.15 - 4.21 $\mu_B$ using the AMF implementation. Finally, when $U$ = 4~eV, the magnetic moment on the Mn atom is between 4.50 - 4.55 $\mu_B$ using the FLL implementation and 4.17 - 4.20 $\mu_B$ using AMF implementation. Overall, the calculated values of the moments are lower using AMF implementation. 

For $U$ = 0~eV, the AFM phase is the lowest energy as predicted by all-electron and pseudopotential calculations.  For $U$ = 3~eV and 4~eV, the FM is the lowest energy phase as predicted by the fully localized limit implementation of LDA+$U$ in Elk and VASP. Additionally, spin-orbit coupling favors antiparallel spin. Surprisingly, the around mean-field approximation in Elk predicts the AFM phase as the lowest energy phase. Overall, $U$ = 4~eV has a strong favor towards parallel spin over $U$ = 3~eV.

%
\begin{figure}[h]
\includegraphics[width=0.8\linewidth, angle=0] {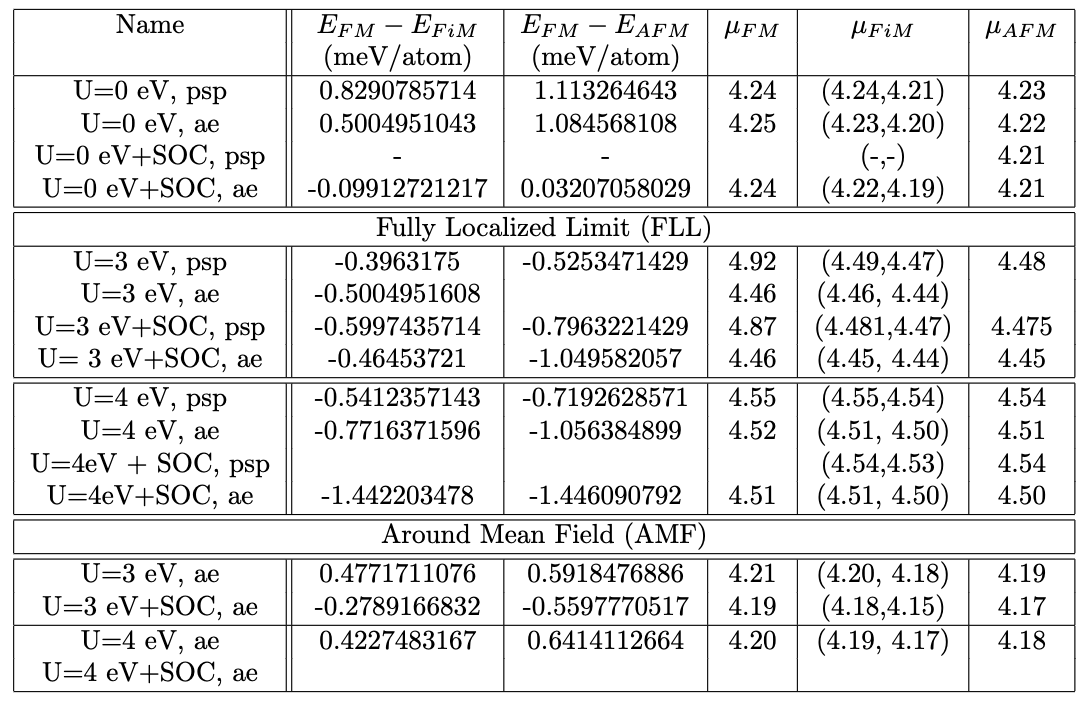}
\caption{Energy differences and magnetic moments on the Mn atom for different values of U. Comparisons with and without SOC have also been made.} 
\label{psp-ae}
\end{figure} 
%

\bibliographystyle{apsrev}
\bibliography{biblio.bib}